\documentclass[12pt]{article}
\usepackage{graphicx}
\renewcommand{\baselinestretch}{1.}
\begin{document}

\begin{center}
{\Large \bf An evidence of mass dependent differential kinetic
freeze-out scenario observed in Pb-Pb collisions at 2.76 TeV}

\vskip1.0cm

Hai-Ling Lao$^{a}$, Hua-Rong Wei$^{a}$, Fu-Hu
Liu$^{a,}${\footnote{E-mail: fuhuliu@163.com;
fuhuliu@sxu.edu.cn}}, and Roy A. Lacey$^{b,}${\footnote{E-mail:
Roy.Lacey@Stonybrook.edu}}

{\small\it $^a$Institute of Theoretical Physics, Shanxi
University, Taiyuan, Shanxi 030006, China

$^b$Departments of Chemistry \& Physics, Stony Brook University,
Stony Brook, NY 11794, USA}
\end{center}

\vskip1.0cm

{\bf Abstract:} Transverse momentum spectra of different particles
produced in mid-rapidity interval in lead-lead (Pb-Pb) collisions
with different centrality intervals, measured by the ALICE
Collaboration at center-of-mass energy per nucleon pair
$\sqrt{s_{NN}}=2.76$ TeV, are conformably and approximately
described by the Tsallis distribution. The dependences of
parameters (effective temperature, entropy index, and
normalization factor) on event centrality and particle rest mass
are obtained. The source temperature at the kinetic freeze-out is
obtained to be the intercept in the linear relation between
effective temperature and particle rest mass, while the particle
(transverse) flow velocity in the source rest frame is extracted
to be the slope in the linear relation between mean (transverse)
momentum and mean moving mass. It is shown that the source
temperature increases with increase of particle rest mass, which
exhibits an evidence of mass dependent differential kinetic
freeze-out scenario or multiple kinetic freeze-out scenario.
\\

{\bf Keywords:} Source temperature, kinetic freeze-out scenario,
mass dependent differential kinetic freeze-out scenario
\\

{\bf PACS:} 12.38.Mh, 25.75.Dw, 24.10.Pa

\vskip1.0cm

{\section{Introduction}}

High energy nucleus-nucleus (heavy ion) collisions at the large
hadron collider (LHC) [1--5] have been providing another excellent
environment and condition of high temperature and density, where
the new state of matter, namely the quark-gluon plasma (QGP)
[6--8], is expected to form and to live for a longer lifetime than
that at the relativistic heavy ion collider (RHIC) [9]. Although
the RHIC is scheduled to run at lower energies which are around
the critical energy of phase transition from hadronic matter to
QGP, the LHC is expected to run at higher energies. Presently, the
LHC has provided three different types of collisions:
proton-proton ($pp$), proton-lead ($p$-Pb), and lead-lead (Pb-Pb)
collisions at different collision energies. The former two are not
expected to form the QGP due to small system, though the
deconfinement of quarks and gluons may appear. The latter one is
expected to form the QGP due to large system and high energy.

It is believed that the QGP is formed in Pb-Pb collisions at the
LHC and in nucleus-nucleus collisions at lower energy till dozens
of GeV at the RHIC [10, 11]. If mesons are produced in the
participant region where violent collision had happened and the
QGP is formed, nuclear fragments such as helium or heavier nuclei
are expected to emit in spectator region where non-violent
evaporation and fragmentation had happened [12--14]. The ALICE
Collaboration [15--18] measured together positive pions $\pi^+$,
positive kaons $K^+$, protons $p$, deuterons $d$, and one of
helium isotopes $^3$He in Pb-Pb collisions with different
centrality intervals at the LHC. It gives us a chance to describe
uniformly different particles. In particular, we are interested in
the uniform description of transverse momentum spectra of $\pi^+$,
$K^+$, $p$, $d$, and $^3$He, so that we can extract the kinetic
freeze-out (KFO) temperature of interacting system (i.e. source
temperature at KFO).

From source temperature at KFO, we can draw a KFO scenario. There
are three different KFO scenarios discussed in literature [3,
19--22]. The single KFO scenario [19] uses one set of parameters
for both the spectra of strange and non-strange particles. The
double KFO scenario [3, 20] uses a set of parameters for the
spectra of strange particles, and another set of parameters for
the spectra of non-strange particles. The multi-KFO scenario [21,
22] uses different sets of parameters for different particles with
different masses. Naturally, the mass dependent differential KFO
scenario [22] belongs to the multi-KFO scenario. It is an open
question which KFO scenario describes correctly. We are interested
in the study of KFO scenario in the present work. As can be seen
from the following sections, our analysis provides an evidence of
mass dependent differential KFO scenario.

To extract source temperature at the KFO, we have to describe
transverse momentum spectra. More than ten functions are used in
the descriptions of transverse momentum spectra. In the present
work, we select the Tsallis distribution [23--25] that covers the
sum of two or three standard distributions [26, 27] and describes
temperature fluctuations among different local equilibrium states.
Based on the descriptions of the experimental data of the ALICE
Collaboration [15, 16] on Pb-Pb collisions at center-of-mass
energy per nucleon pair $\sqrt{s_{NN}}=2.76$ TeV, the source
temperature at the KFO is obtained to be the intercept in the
linear relation between effective temperature and particle rest
mass, while the particle (transverse) flow velocity in the source
rest frame is extracted to be the slope in the linear relation
between mean (transverse) momentum and mean moving mass. If we use
other functions, the method is in fact the same. Because of no
difference between positive and negative spectra being reported
[16], we are just fitting the available positive data in the
analysis.

The structure of the present work is as followings. The model and
method are shortly described in section 2. Results and discussion
are given in section 3. In section 4, we summarize our main
observations and conclusions.
\\

{\section{The model and method}}

We discuss the collision process in the framework of the
multisource thermal model [28--30]. According to the model, many
emission sources are formed in high energy nucleus-nucleus
collisions. We can choose different distributions to describe the
emission sources and particle spectra. These distributions
include, but are not limited to, the Tsallis distribution
[23--25], the standard (Boltzmann, Fermi-Dirac, and Bose-Einstein)
distributions [26], the Tsallis + standard distributions [31--36],
the Erlang distribution [28], and so forth.

The Tsallis distribution can be described by two or three standard
distributions. The Tsallis + standard distributions can be
described by two or three Tsallis distributions [27]. It is
needless to choose the standard distributions due to multiple
sources (temperatures). It is also needless to choose the Tsallis
+ standard distributions due to not too many sources
(temperatures). A middle way is to choose the Tsallis distribution
which describes the temperature fluctuation in a few sources to
give an average value. These sources with different excitation
degrees can be naturally described by the standard distributions
with different effective temperatures, which result from the
multisource thermal model [28--30].

The Tsallis distribution has more than one function forms [23--25,
31--38]. In the rest frame of a considered source, we choose a
simplified form of the joint probability density function of
transverse momentum ($p_T$) and rapidity ($y$),
\begin{equation}
f(p_T,y) \propto \frac{d^2N}{dydp_T}=\frac{gV}{(2\pi)^2} p_T
\sqrt{p^2_T+m^2_0}\cosh y \bigg[ 1+\frac{q-1}{T} \Big(
\sqrt{p^2_T+m^2_0}\cosh y-\mu \Big) \bigg]^{-q/(q-1)},
\end{equation}
where $N$ is the particle number, $g$ is the degeneracy factor,
$V$ is the volume of emission sources, $T$ is the temperature
which describes averagely a few sources (local equilibrium
states), $q$ is the entropy index which describes the degree of
non-equilibrium among different states, $\mu$ is the chemical
potential which is related to $\sqrt{s_{NN}}$ [39] and can be
regarded as 0 at the LHC, $m_0$ is the rest mass of the considered
particle. Generally, the 4-parameter ($T$, $q$, $\mu$, and $V$)
form of Eq. (1) is capable of reproducing the particle spectra,
where $T$, $q$, and $\mu$ are fitted independently for the
considered particle species, and $V$ is related to other
parameters.

Eq. (1) results in the transverse momentum probability density
function which is an alternative representation of the Tsallis
distribution as follows
\begin{equation}
f_{p_T}(p_T)=\frac{1}{N} \frac{dN}{dp_T} =
\int^{y_{\max}}_{y_{\min}} f(p_T,y) dy,
\end{equation}
where $y_{\max}$ and $y_{\min}$ denote the maximum and minimum
rapidities, respectively. Similarly, Eq. (1) results in the
rapidity probability density function in the source rest frame as
follows
\begin{equation}
f_y(y)=\frac{1}{N} \frac{dN}{dy} = \int^{p_{T\max}}_0 f(p_T,y)
dp_T,
\end{equation}
where $p_{T\max}$ denotes the maximum transverse momentum.

Under the assumption of isotropic emission in the source rest
frame, we have the polar angle probability density function to be
\begin{equation}
f_{\theta}(\theta)=\frac{1}{2} \sin \theta.
\end{equation}
Let $r_1$ and $r_2$ denote the random numbers distributed
uniformly in [0,1] respectively. We can use the Monte Carlo method
to obtain a series of $p_T$ which satisfies
\begin{equation}
\int_0^{p_T} f_{p_T}(p_T) dp_T <r_1< \int_0^{p_T+dp_T}
f_{p_T}(p_T) dp_T.
\end{equation}
The Monte Carlo method results in
\begin{equation}
\theta=2\arcsin \sqrt{r_2}
\end{equation}
due to Eq. (4). Thus, we can obtain a series of values of momentum
and energy due to the momentum $p=p_T/\sin \theta$ and the energy
$E=\sqrt{p^2+m_0^2}$. The energy $E$ is in fact equal to the
moving mass $m$ in the natural unit system. Then, we have the mean
moving mass $\overline{m}$ to be the mean energy $\overline{E}$.
\\

{\section{Results and discussion}}

\begin{figure}
\hskip-1.0cm \begin{center}
\includegraphics[width=12.0cm]{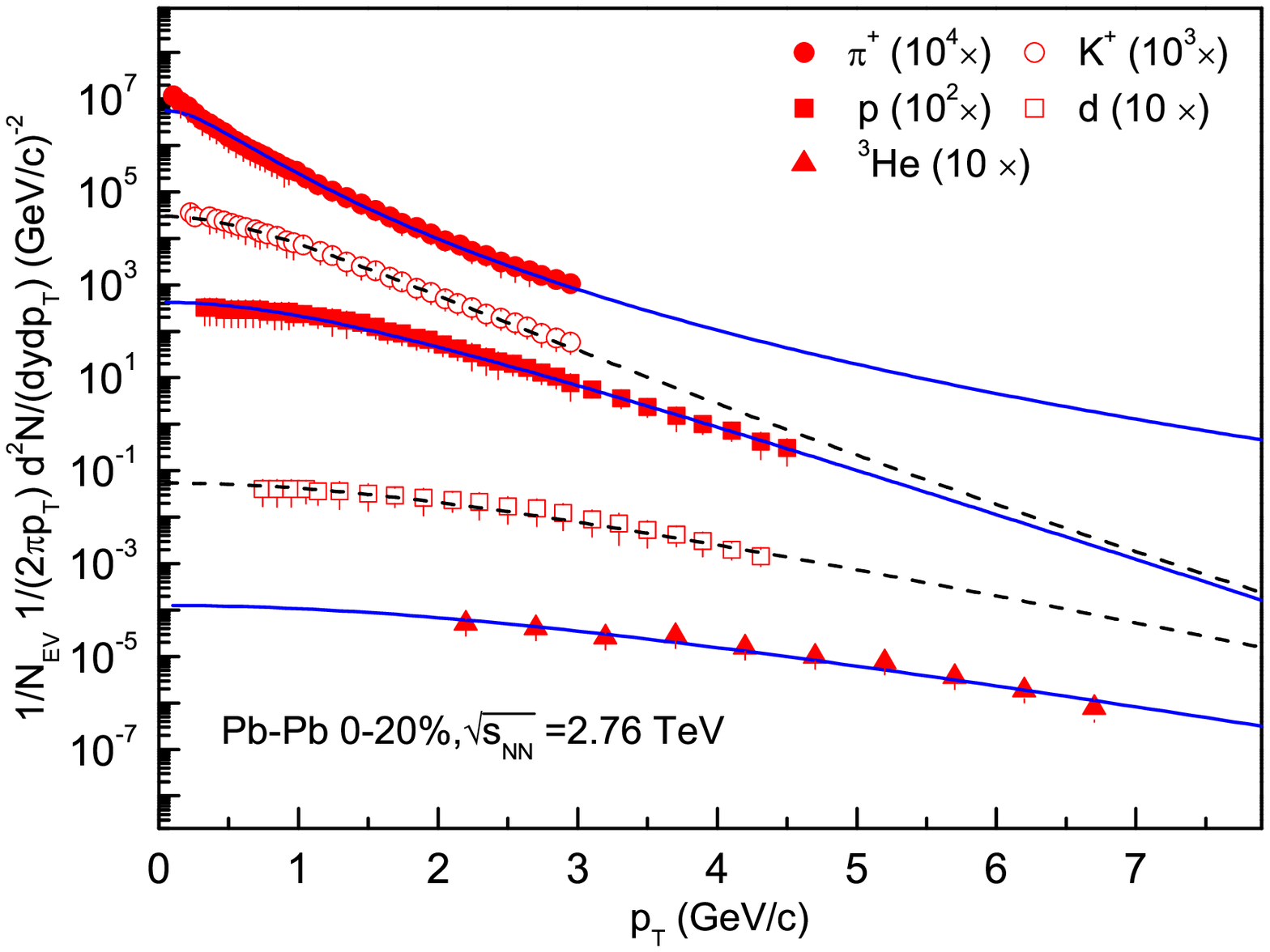}
\end{center}
\vskip.0cm Figure 1. Transverse momentum spectra of $\pi^+$,
$K^+$, $p$, $d$, and $^3$He produced in mid-rapidity interval
($|y|<0.5$) in Pb-Pb collisions at $\sqrt{s_{NN}}=2.76$ TeV. The
symbols represent the experimental data of the ALICE Collaboration
[16] in centrality interval 0--20\%, which are scaled by different
amounts marked in the panel. The curves are our results fitted by
using the Tsallis distribution based on Eq. (1). In the fitting,
the method of least squares is used to obtain the values of
related parameters.
\end{figure}

Figure 1 presents the transverse momentum spectra, $(1/N_{EV})
d^2N/(2\pi p_T dy dp_T)$, of $\pi^+$, $K^+$, $p$, $d$, and $^3$He
produced in mid-rapidity interval ($|y|<0.5$) in Pb-Pb collisions
at $\sqrt{s_{NN}}=2.76$ TeV, where $N_{EV}$ denotes the number of
events. The symbols represent the experimental data of the ALICE
Collaboration [16] in centrality interval 0--20\%, which are
scaled by different amounts marked in the panel. The curves are
our results fitted by using the Tsallis distribution based on Eq.
(1) at mid-rapidity ($y=0$). In the fitting, the method of least
squares is used. The values of related parameters $T$, $q$, and
$N_0$ are listed in Table 1 with values of $\chi^2$ per degree of
freedom ($\chi^2$/dof), where
\begin{equation}
N_0=\frac{gV}{(2\pi)^3} \int_{0}^{p_{T\max}}
\int_{y_{\min}}^{y_{\max}} \sqrt{p^2_T+m^2_0}\cosh y \bigg[
1+\frac{q-1}{T} \Big( \sqrt{p^2_T+m^2_0}\cosh y-\mu \Big)
\bigg]^{-q/(q-1)} dy dp_T
\end{equation}
is the normalization factor which is used to compare the
normalized curve with experimental data. One can see that the
Tsallis distribution describes conformably and approximately
$\pi^+$, $K^+$, $p$, $d$, and $^3$He spectra. The effective
temperature increases with increase of particle rest mass.
\\
\\

\renewcommand{\baselinestretch}{0.8}
{\small {Table 1. Values of $T$, $q$, $N_0$, and $\chi^2$/dof
corresponding to the curves in Figures 1--3, where the data set
for $^3$He in Figure 1 is the same as that for centrality interval
0--20\% in Figure 3 [16], which renders the same values of
parameters.
{%
\begin{center}
\begin{tabular}{ccccccc}
\hline\hline  Figure & Type 1 & Type 2 & $T$ (GeV) & $q$  & $N_0$ & $\chi^2$/dof \\
\hline
Figure 1 & 0--20\% & $\pi^+$  & $0.134\pm0.012$ & $1.0980\pm0.0100$ & $228.512\pm81.588$             & 0.163 \\
         &         & $K^+$    & $0.271\pm0.020$ & $1.0210\pm0.0060$ & $22.238\pm6.254$               & 0.082 \\
         &         & $p$      & $0.411\pm0.033$ & $1.0010\pm0.0009$ & $4.709\pm1.502$                & 0.077 \\
         &         & $d$      & $0.645\pm0.068$ & $1.0010\pm0.0008$ & $0.010\pm0.003$                & 0.117 \\
         &         & $^{3}$He & $0.781\pm0.085$ & $1.0013\pm0.0009$ & $(2.930\pm0.453)\times10^{-5}$ & 1.162 \\
\hline
Figure 2 & $d$     & 0--10\%  & $0.659\pm0.039$ & $1.0005\pm0.0004$ & $(1.184\pm0.105)\times10^{-2}$ & 2.359 \\
         &         & 10--20\% & $0.634\pm0.028$ & $1.0004\pm0.0003$ & $(9.469\pm0.855)\times10^{-3}$ & 1.689 \\
         &         & 20--40\% & $0.582\pm0.031$ & $1.0007\pm0.0006$ & $(6.256\pm0.742)\times10^{-3}$ & 0.863 \\
         &         & 40--60\% & $0.485\pm0.056$ & $1.0010\pm0.0008$ & $(2.937\pm0.612)\times10^{-3}$ & 0.244 \\
         &         & 60--80\% & $0.345\pm0.021$ & $1.0010\pm0.0008$ & $(9.317\pm1.682)\times10^{-4}$ & 0.214 \\
         &         & $pp$     & $0.273\pm0.014$ & $1.0030\pm0.0010$ & $(4.143\pm0.545)\times10^{-5}$ & 0.831 \\
\hline
Figure 3 & $^{3}$He& 0--20\%  & $0.781\pm0.085$ & $1.0013\pm0.0009$ & $(2.930\pm0.453)\times10^{-5}$ & 1.162 \\
         &         & 20--80\% & $0.670\pm0.032$ & $1.0021\pm0.0015$ & $(6.100\pm0.951)\times10^{-6}$ & 1.379 \\
\hline \hline
\end{tabular}%
\end{center}
}} }
\renewcommand{\baselinestretch}{1.0}
\vskip1.0cm

\begin{figure}
\hskip-1.0cm \begin{center}
\includegraphics[width=12.0cm]{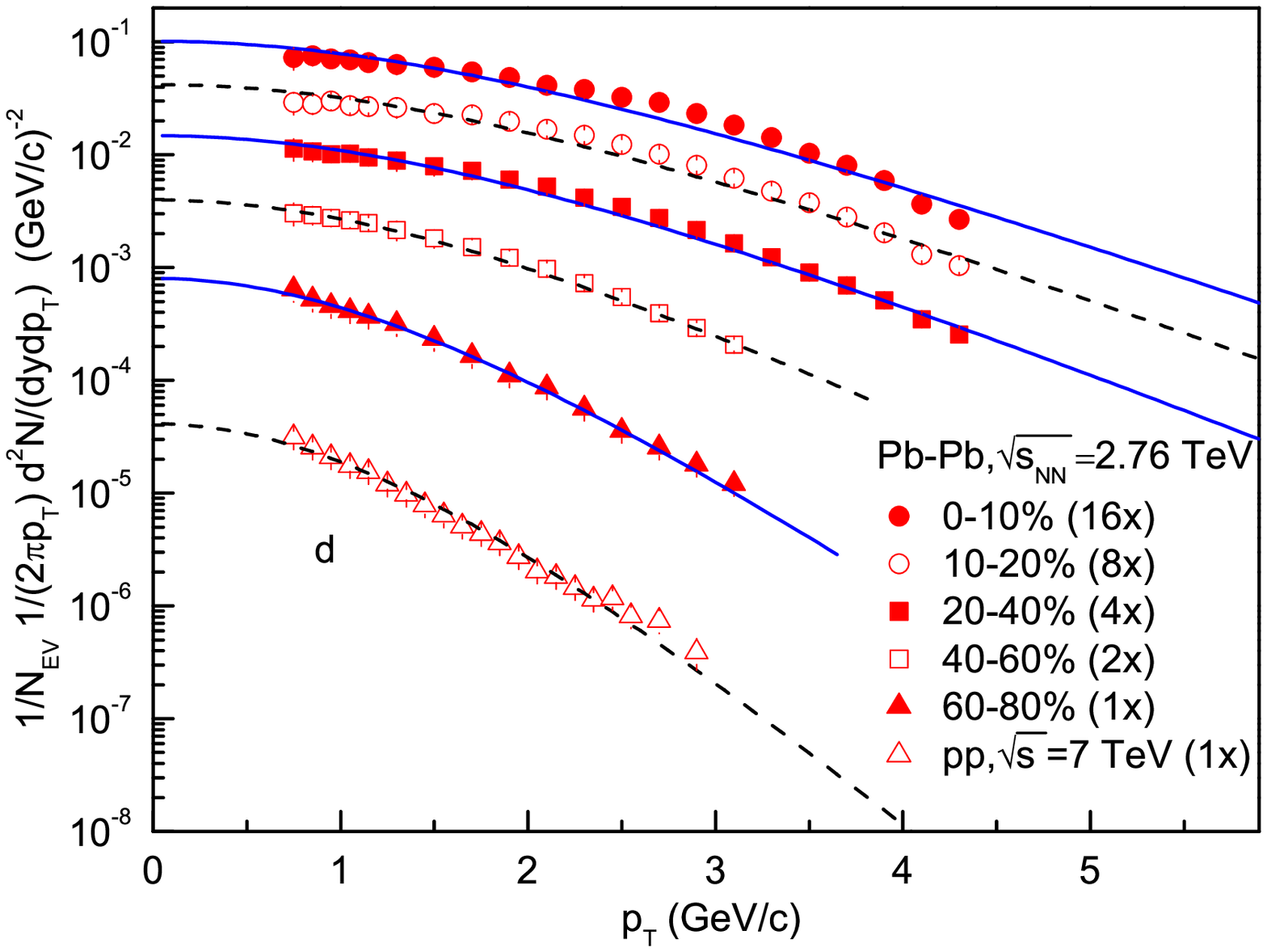}
\end{center}
\vskip.0cm Figure 2. The same as for Figure 1, but showing the
results of $d$ in Pb-Pb collisions with different centrality
intervals and in $pp$ collisions.
\end{figure}

\begin{figure}
\hskip-1.0cm \begin{center}
\includegraphics[width=12.0cm]{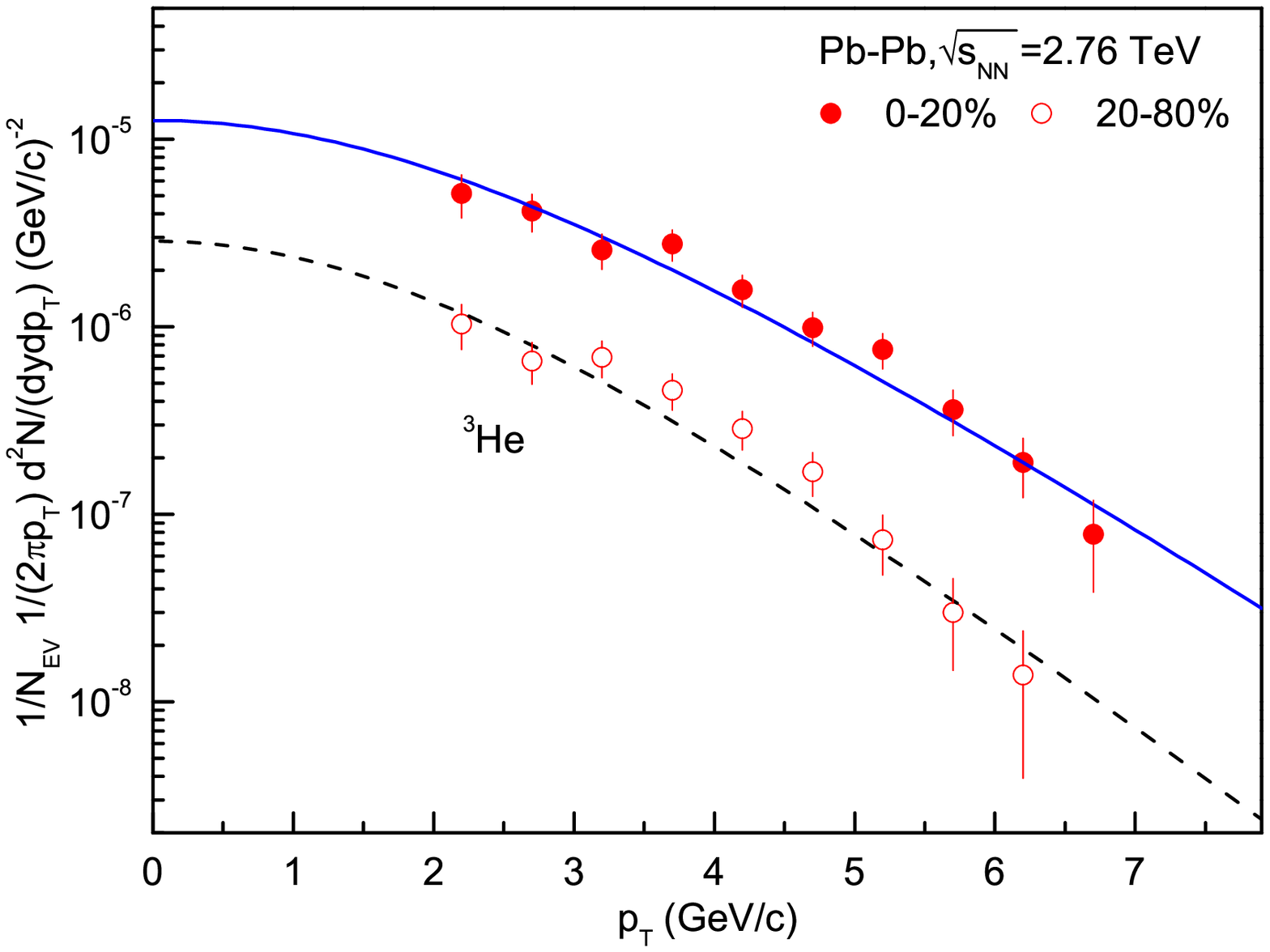}
\end{center}
\vskip.0cm Figure 3. The same as for Figure 1, but showing the
results of $^3$He in Pb-Pb collisions with two centrality
intervals.
\end{figure}

Figure 2 is similar to Figure 1, but it shows the results for $d$
in different centrality intervals, which are scaled by different
amounts marked in the panels. At the same time, the result in $pp$
collisions at $\sqrt{s}=7$ TeV is presented for comparison, where
$\sqrt{s}$ is a simplified form of $\sqrt{s_{NN}}$ for $pp$
collisions. Figure 3 is similar to Figure 1, but it shows the
results for $^3$He in centrality intervals 0--20\% and 20--80\%,
where the data set for 0--20\% is the same as that for $^3$He in
Figure 1 [16]. The related parameter values are listed in Table 1
with values of $\chi^2$/dof. One can see that the Tsallis
distribution describes approximately the experimental data of $d$
produced in Pb-Pb collisions with different centrality intervals
at the LHC. The effective temperature extracted from $d$ spectra
decreases with decrease of centrality (or increase of centrality
percentage).

To study the change trends of parameters with centrality interval
($C$) of event and rest mass of particle, Figure 4 gives the
dependences of (a) $T$ on $C$ for $d$ in events with different
centrality intervals and (b) $T$ on $m_0$ for particles in events
with centrality interval 0--20\%, where only the result for $d$ in
Figure 4(a) is available due to the experimental result [16]. The
symbols represent the parameter values extracted from Figures 1
and 2 and listed in Table 1, and the curves are our results fitted
by the method of least squares. The curve in Figure 4(a) is
described by
\begin{equation}
T=-(0.000059\pm0.000005)C^2-(0.0012\pm0.0002)C+(0.668\pm0.005)
\end{equation}
with $\chi^2$/dof=0.927, where $T$ is in the units of GeV. The
solid, dotted, and dashed curves in Figure 4(b) are linear
fittings for i) $\pi^+$, $K^+$, and $p$; ii) $\pi^+$, $K^+$, $p$,
and $d$; and iii) $\pi^+$, $K^+$, $p$, $d$, and $^3$He, which are
described by
\begin{equation}
T=(0.091\pm0.009)+(0.345\pm0.014)m_0,
\end{equation}
\begin{equation}
T=(0.115\pm0.017)+(0.291\pm0.016)m_0,
\end{equation}
and
\begin{equation}
T=(0.148\pm0.032)+(0.241\pm0.020)m_0,
\end{equation}
with $\chi^2$/dof=0.429, 2.091, and 5.869, respectively, where
$m_0$ is in the units of GeV/$c^2$.

The intercept in Eq. (9) is regarded as the KFO temperatures
[40--43] of emission source, which is 0.091 GeV corresponding to
massless particles, when the source produces $\pi^+$, $K^+$, and
$p$. Including $d$ causes a large intercept (0.115 GeV) in Eq.
(10), while including $d$ and $^3$He causes a larger intercept
(0.148 GeV) in Eq. (11). Although the errors in intercepts are
large, these results render that the KFO temperature increases
with increase of particle rest mass. This is an evidence of mass
dependent differential KFO scenario or multiple KFO scenario [21,
22].

The blast-wave model [44] gives the KFO temperature extracted from
$d$ spectra being 0.077--0.124 GeV and from $^3$He spectra being
0.101 GeV [16] which are comparable with the present work. In
particular, the blast-wave model gives the KFO temperature in
central collisions being less than that in peripheral collisions
[16], which is inconsistent with Figure 4(a) which shows an
opposite result on correlation between effective temperature and
centrality. Although the result of blast-wave model can be
explained as that the interacting system in central collisions
undergoes a longer kinetic evolution which results in a lower KFO
temperature comparing with peripheral collisions, the present
result can be explained as that the interacting system in central
collisions stays in a higher excitation state comparing with
peripheral collisions.

On the other hand, we have used an alternative method to extract
indirectly the KFO temperature based on the linear relation
between effective temperature and rest mass [40--43]. The evidence
coming from similar analyses in RHIC and LHC experiments [45, 46],
where the fit parameters have been studied also against
centrality, even down to data of $d$-nucleus or $pp$ collisions,
confirms that the KFO temperature in central collisions is less
than that in peripheral collisions, which is inconsistent with the
present work. For central and peripheral collisions, the relative
size of KFO temperature obtained in the present work is similar to
those of the chemical freeze-out temperature and effective
temperature. We would like to point out that the present work is
qualitatively consistent with ref. [19], where the Tsallis +
blast-wave model is used at RHIC energy.

Although the interpretation of the Tsallis distribution is still
controversial, at least in the field of concern here, it could be
interesting to learn the behavior of non-additive entropy. In
Figure 5, the dependences of (a) $q$ on $C$ for $d$ in events with
different centrality intervals and (b) $q$ on $m_0$ for different
particles in events with centrality interval 0--20\% are given,
where only the result for $d$ in Figure 5(a) is available due to
the experimental result [16]. The symbols represent the parameter
values extracted from Figures 1 and 2 and listed in Table 1. For
$d$ in events with different centrality intervals, the values of
$q$ are foregone to be consistent with one [Figure 5(a)]. For the
events with centrality interval 0--20\%, $\pi^+$ corresponds to a
larger $q$ than others [Figure 5(b)]. This renders that the
production of pions is more polygenetic than others. Because of
the most values of $q$ being small, the interacting system stays
approximately in an equilibrium state.

In fact, the ranges of most $p_T$ spectra considered in Pb-Pb
collisions in the present work are narrow, which result mainly
from the soft process which is a single source and can be
described by the standard distribution. If we study wide $p_T$
spectra, both the soft and hard processes have to be considered.
We need two or three standard distributions, the standard
distribution + a power law, or the Tsallis distribution with large
$q$ to describe the wide spectra. The situation for $pp$
collisions is similar to Pb-Pb collisions. The advantage of
Tsallis distribution will appear in description of the wide
spectra. For the narrow spectra, both the standard distribution
and the Tsallis distribution with small $q$ are satisfied.
However, we use the Tsallis distribution due to its potential
application in wide $p_T$ spectra. In addition, the STAR
experiment already tried a Tsallis-like study, publishing also a
Tsallis + blast-wave model-based interpretation of their data [9].
The ALICE data, on the other hand, have been compared to a
blast-wave + thermal-based fit [16]. In both cases the agreement
is very good. These facts render that the Tsallis distribution has
a wide application in high energy physics.

In Figure 6, the dependences of (a) $N_0$ on $C$ for $d$ in events
with different centrality intervals and (b) $N_0$ on $m_0$ for
different particles in events with centrality interval 0--20\% are
given, where only the result for $d$ in Figure 6(a) is available
due to the experimental result [16]. The symbols represent the
parameter values extracted from Figures 1 and 2 and listed in
Table 1. The curves are our results fitted by the method of least
squares, which are described by
\begin{equation}
N_0=(0.0166\pm0.0004)\exp[-(0.019\pm0.001)C]-(0.0033\pm0.0002)
\end{equation}
and
\begin{equation}
N_0=(580.207\pm62.425)\exp[-(5.912\pm0.118)m_0]
\end{equation}
with $\chi^2$/dof=0.572 and 1.712 respectively. It is shown that
$N_0$ decreases with decrease of centrality. The larger the
particle rest mass is, the lower the production probability is.
Although $N_0$ is only a normalization factor and the data are not
cross-section, they are proportional to the volumes of sources
producing different particles. Therefore, studying $N_0$
dependence is significative.

\begin{figure}
\hskip-1.0cm \begin{center}
\includegraphics[width=15.0cm]{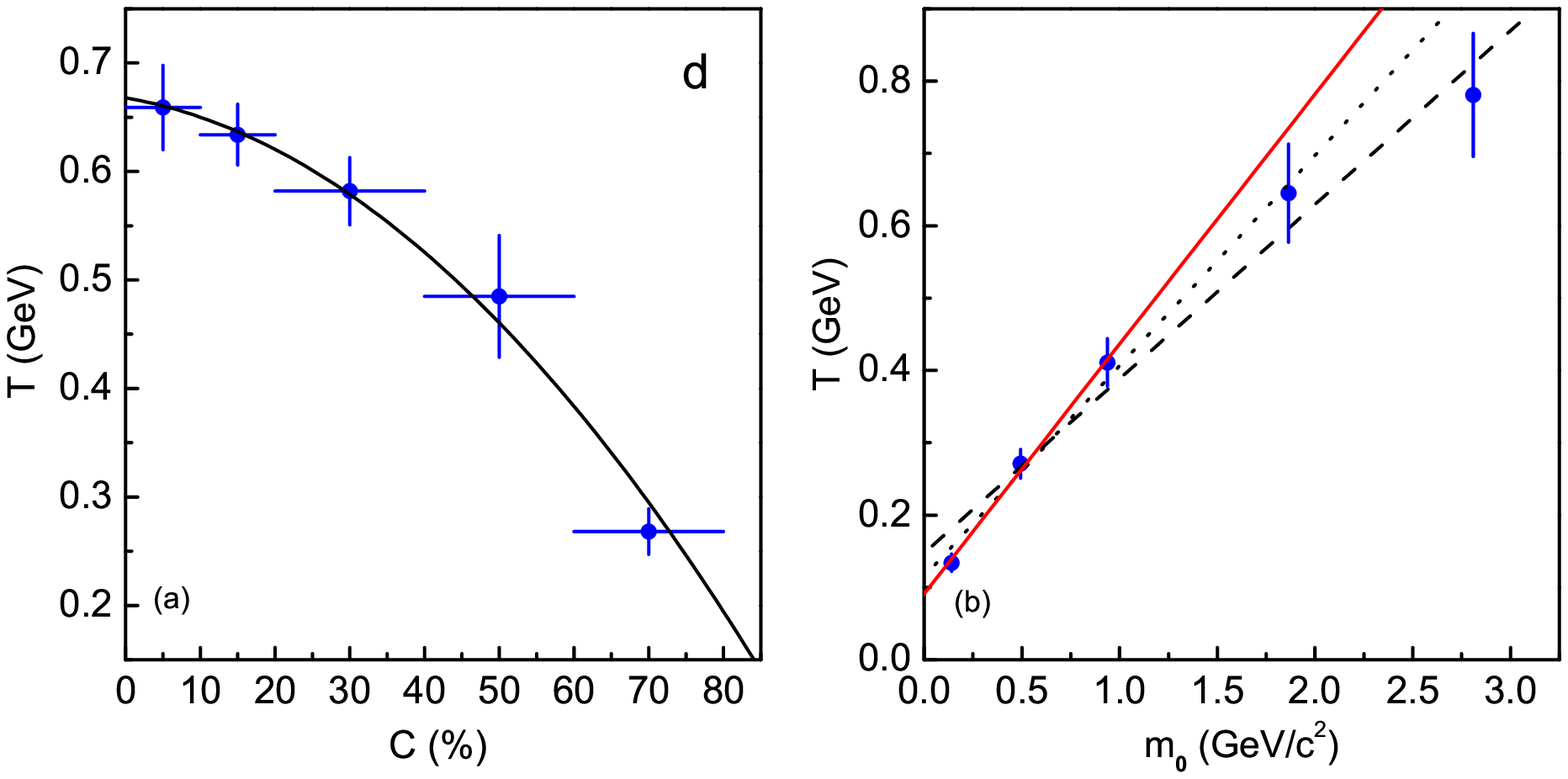}
\end{center}
\vskip.0cm Figure 4. Dependences of (a) $T$ on $C$ for $d$ in
events with different centrality intervals and (b) $T$ on $m_0$
for particles in events with centrality interval 0--20\%. The
symbols represent the parameter values listed in Table 1. The
curves and lines are our results fitted by the method of least
squares.
\end{figure}

\begin{figure}
\hskip-1.0cm \begin{center}
\includegraphics[width=15.0cm]{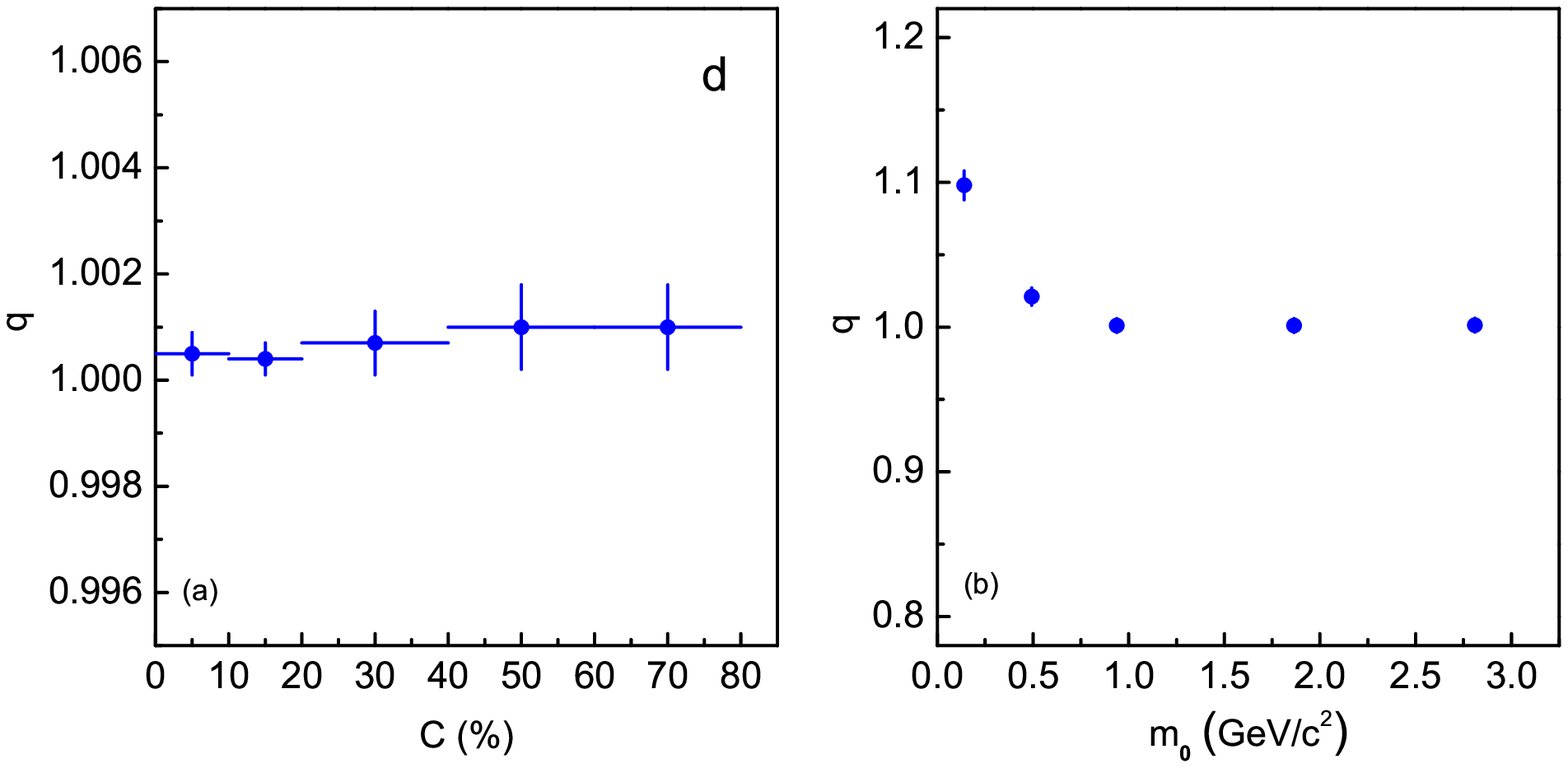}
\end{center}
\vskip.0cm Figure 5. Dependences of (a) $q$ on $C$ for $d$ in
events with different centrality intervals and (b) $q$ on $m_0$
for particles in events with centrality interval 0--20\%. The
symbols represent the parameter values listed in Table 1.
\end{figure}

\begin{figure}
\hskip-1.0cm \begin{center}
\includegraphics[width=15.0cm]{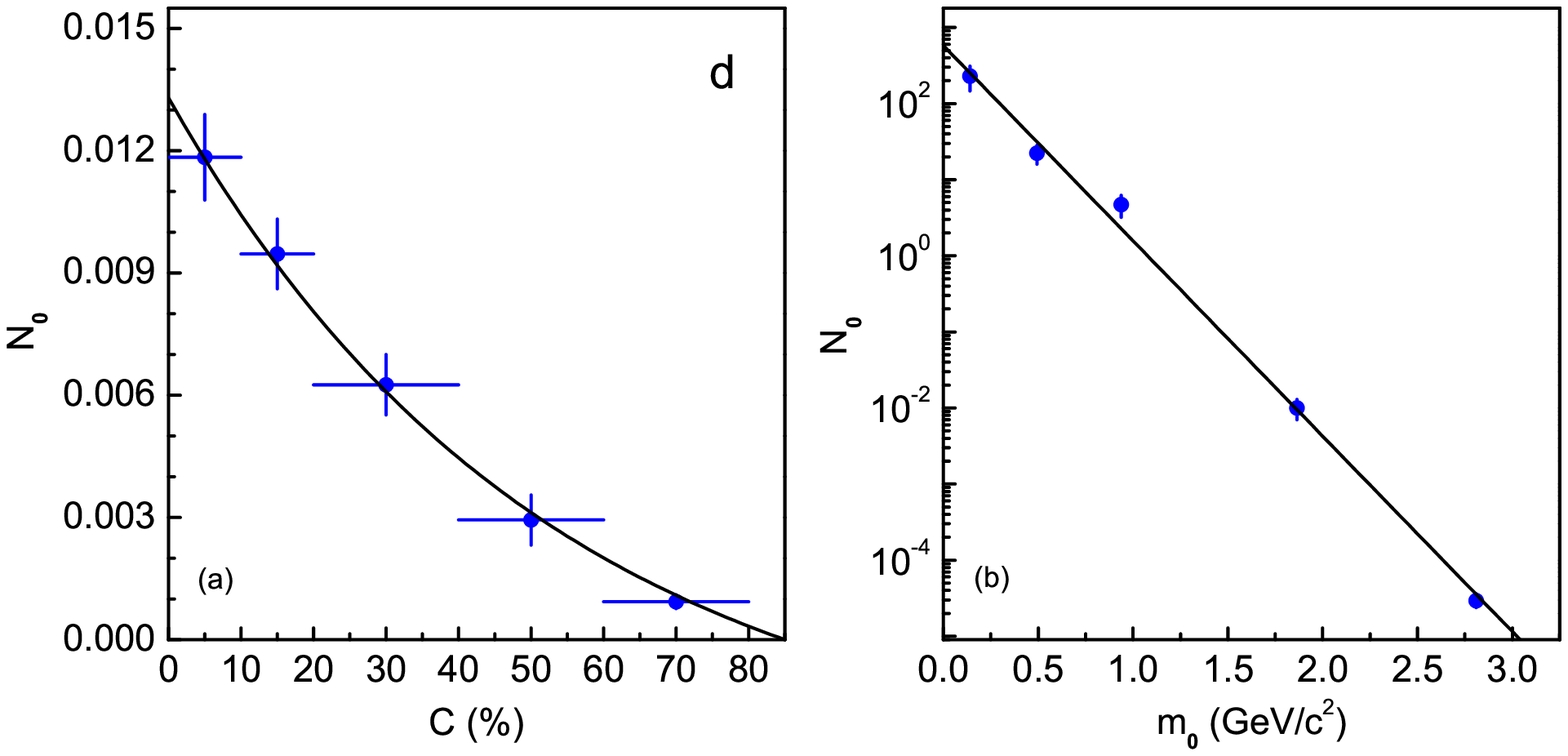}
\end{center}
\vskip.0cm Figure 6. Dependences of (a) $N_0$ on $C$ for $d$ in
events with different centrality intervals and (b) $N_0$ on $m_0$
for particles in events with centrality interval 0--20\%. The
symbols represent the parameter values listed in Table 1. The
curve and line are our results fitted by the method of least
squares.
\end{figure}

\begin{figure}
\hskip-1.0cm
\begin{center}
\includegraphics[width=15.0cm]{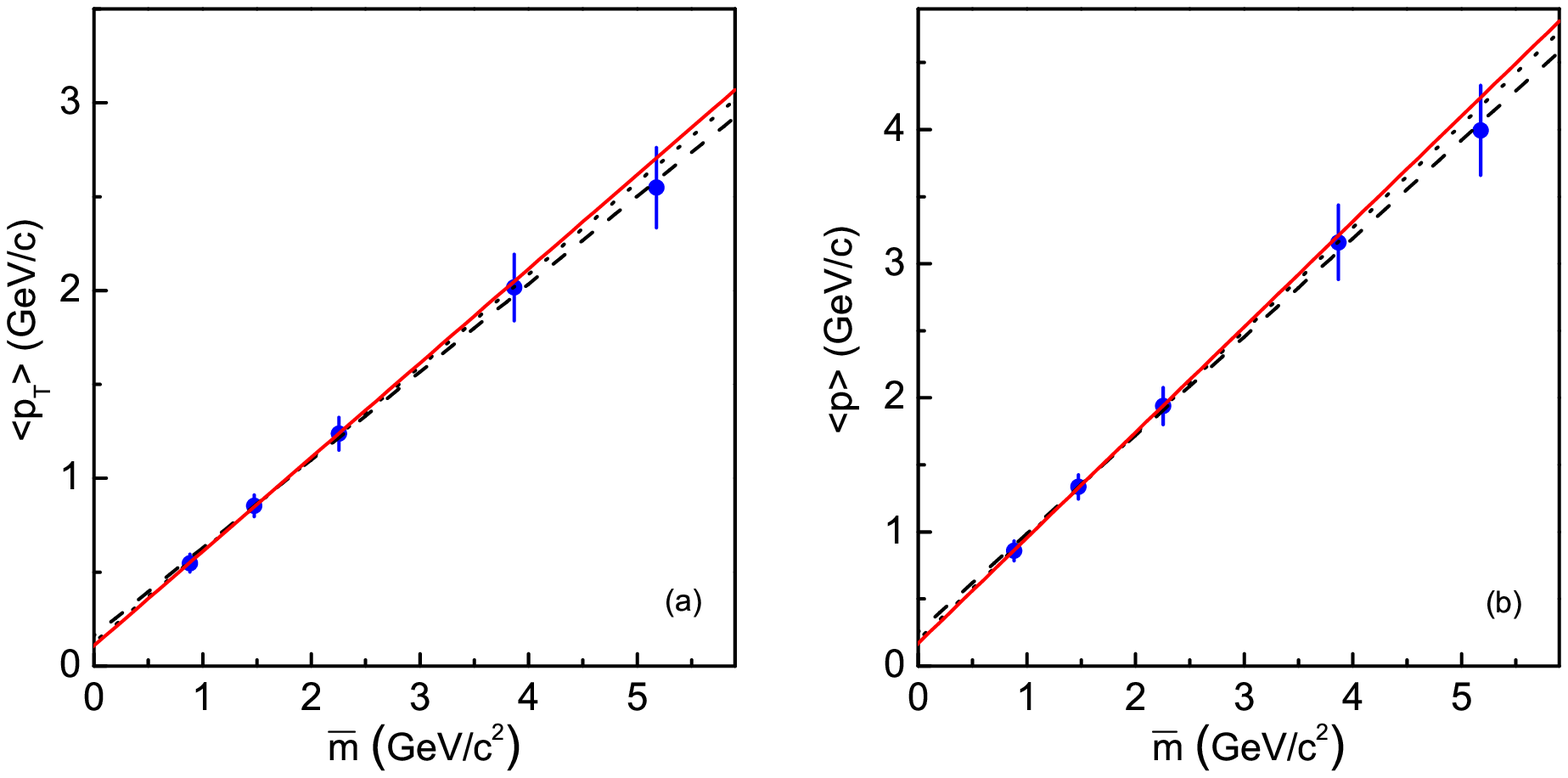}
\end{center}
\vskip.0cm Figure 7. Dependences of (a) $\langle p_T \rangle$ on
$\overline{m}$ and (b) $\langle p \rangle$ on $\overline{m}$ for
particles in events with centrality interval 0--20\%. The symbols
represent the values of $\langle p_T \rangle$ and $\langle p
\rangle$ at different $\overline{m}$, which are calculated by
using the Monte Carlo method in the source rest frame. The lines
are our results fitted by the method of least squares.
\end{figure}

To extract the transverse flow velocity, we display the dependence
of mean transverse momentum ($\langle p_T \rangle$) on mean moving
mass ($\overline{m}$) in Figure 7(a). The symbols represent the
values of $\langle p_T \rangle$ and $\overline{m}$ for different
particles calculated by using the Monte Carlo method in the source
rest frame. The solid, dotted, and dashed curves in Figure 7(a)
are linear fittings for i) $\pi^+$, $K^+$, and $p$; ii) $\pi^+$,
$K^+$, $p$, and $d$; and iii) $\pi^+$, $K^+$, $p$, $d$, and
$^3$He, which are described by
\begin{equation}
\langle p_T \rangle=(0.108\pm0.006)+(0.502\pm0.004)\overline{m},
\end{equation}
\begin{equation}
\langle p_T \rangle=(0.124\pm0.008)+(0.490\pm0.003)\overline{m},
\end{equation}
and
\begin{equation}
\langle p_T \rangle=(0.162\pm0.029)+(0.468\pm0.009)\overline{m},
\end{equation}
with $\chi^2$/dof=0.009, 0.025, and 0.155, respectively, where
$\langle p_T \rangle$ and $\overline{m}$ are in the units of
GeV/$c$ and GeV/$c^2$ respectively. From the consideration of
dimension, the slopes in Eqs. (14)--(16) are regarded as the
(average) transverse flow velocity, which is close to $0.5c$.
Including $d$ or $d$ and $^3$He, one can see a small decrease in
transverse flow velocity. The blast-wave model [44] gives the
transverse flow velocity for $d$ is 0.38--0.63$c$ and for $^3$He
is 0.56--0.57$c$ [16] which is comparable with the present work.

Figure 7(b) is the same as for Figure 7(a), but showing the
dependence of mean momentum ($\langle p \rangle$) on
$\overline{m}$. The values of $\langle p \rangle$ are calculated
by using the Monte Carlo method in the source rest frame, too. The
solid, dotted, and dashed curves in Figure 7(b) are linear
fittings for i) $\pi^+$, $K^+$, and $p$; ii) $\pi^+$, $K^+$, $p$,
and $d$; and iii) $\pi^+$, $K^+$, $p$, $d$, and $^3$He, which are
described by
\begin{equation}
\langle p \rangle=(0.170\pm0.010)+(0.786\pm0.006)\overline{m},
\end{equation}
\begin{equation}
\langle p \rangle=(0.195\pm0.013)+(0.768\pm0.005)\overline{m},
\end{equation}
and
\begin{equation}
\langle p \rangle=(0.254\pm0.046)+(0.733\pm0.015)\overline{m},
\end{equation}
with $\chi^2$/dof=0.009, 0.024, and 0.156, respectively, where
$\langle p \rangle$ is in the units of GeV/$c$. From the
consideration of dimension, the slopes in Eqs. (17)--(19) are
regarded as the (average) flow velocity, which is close to
$(\pi/2)0.5c$ which confirms our very recent work [47]. Including
$d$ or $d$ and $^3$He, one can see a small decrease in flow
velocity.

In the above discussions, although KFO temperatures and
(transverse) flow velocities are only extracted by us by indirect
methods and we have not obtained the straightforward KFO, the
present work provides anyhow alternative methods to describe $p_T$
spectra and to extract indirectly KFO temperature and (transverse)
flow velocity. In fact, different functions result in different
KFO temperatures. To extract the absolute temperature at KFO, the
standard distribution is the best choice. However, we have to use
a multi-component standard distribution. Because the mean
(transverse) momentum and mean moving mass are independent of
models. The (transverse) flow velocity extracted from the slope in
the linear relation between mean (transverse) momentum and mean
moving mass should be independent of models, too.

In our very recent work [47], the linear relations between $T$ and
$m_0$, $T$ and $\overline{m}$, $\langle p_T \rangle$ and $m_0$,
$\langle p_T \rangle$ and $\overline{m}$, $\langle p \rangle$ and
$m_0$, as well as $\langle p \rangle$ and $\overline{m}$ are
studied. It is shown that the intercept in the linear relation
between $T$ and $m_0$ can be regarded as the KFO temperature, the
slope in the linear relation between $\langle p_T \rangle$ and
$\overline{m}$ can be regarded as the transverse flow velocity,
and the slope in the linear relation between $\langle p \rangle$
and $\overline{m}$ can be regarded as the flow velocity. In the
present work, we use the same treatment to obtain the KFO
temperature, transverse flow velocity, and flow velocity in the
source rest frame. The present work shows that light particles
correspond to low KFO temperature, which reflects that light
particles freeze later than heavy particles. At the same time, due
to small mass, light particles have larger (transverse) flow
velocity than heavy particles.
\\

{\section{Conclusions}}

We summarize here our main observations and conclusions.

a) The transverse momentum distributions of $\pi^+$, $K^+$, $p$,
$d$, and $^3$He produced in Pb-Pb collisions at 2.76 TeV with
different centrality intervals are conformably analyzed by using
the Tsallis distribution. The results calculated by us can fit
approximately the experimental data of the ALICE Collaboration.
The values of parameters such as the effective temperature,
entropy index, and normalization factor are obtained. Small sizes
of entropy index show that the interacting system considered in
the present work stays approximately in an equilibrium state. The
effective temperature extracted from transverse momentum spectra
increases with increase of particle rest mass, and decreases with
decrease of centrality.

b) We have used an alternative method to extract the kinetic
freeze-out temperature of the interacting system based on the
linear relation between the effective temperature and particle
rest mass. The kinetic freeze-out temperature is regarded as the
intercept in the linear relation by us, which shows the same or
similar tendency as those of effective temperature and chemical
freeze-out temperature in the case of studying their dependences
on centrality. The values in central collisions are larger than
those in peripheral collisions, which renders that the interacting
system in central collisions stays in a higher excitation state
comparing with peripheral collisions at kinetic (or chemical)
freeze-out.

c) The kinetic freeze-out temperature extracted by us in 0--20\%
Pb-Pb collisions at 2.76 TeV for including $\pi^+$, $K^+$, and $p$
is 0.091 GeV. Including $d$ causes the kinetic freeze-out
temperature increasing to 0.115 GeV, while including $d$ and
$^3$He causes the kinetic freeze-out temperature increasing to
0.148 GeV. The particle mass effects of kinetic freeze-out
temperature for $d$ and $^3$He are obvious. We think that we have
observed an evidence of mass dependent differential kinetic
freeze-out scenario or multiple kinetic freeze-out scenario.

d) We have also used an alternative method to extract the
transverse flow velocity and flow velocity of the produced
particles in the source rest frame based on the slopes in the
linear relation between the mean transverse momentum and mean
moving mass, as well as the mean momentum and mean moving mass.
The particle mass effects of (transverse) flow velocity for $d$
and $^3$He are not obvious, though light particles have larger
(transverse) flow velocity. The transverse flow velocity and flow
velocity obtained in the present work is close to $0.5c$ and
$(\pi/2)0.5c$ respectively.
\\

{\bf Conflict of Interests}

The authors declare that there is no conflict of interests
regarding the publication of this paper.
\\

{\bf Acknowledgments}

This work was supported by the National Natural Science Foundation
of China under Grant No. 11575103 and the US DOE under contract
DE-FG02-87ER40331.A008.

\end{document}